# Complex Behavior of Stock Markets:
# Processes of Synchronization and Desynchronization during Crises

***Tanya Araújo*[*] and Francisco Louçã*[§]**

Departamento de Economia, ISEG, Technical University of Lisbon
Research Unit on Complexity in Economics (UECE)
[*] tanya@iseg.utl.pt [§] flouc@iseg.utl.pt

Abstract

*This paper investigates the dynamics of in the S&P500 index from daily returns for the last 30 years. Using a stochastic geometry technique, each S&P500 yearly batch of data is embedded in a subspace that can be accurately described by a reduced number of dimensions. Such feature is understood as empirical evidence for the presence of a certain amount of structure in the market. As part of the inquiry into the structure of the market we investigate changes in its volume and shape, and we define new measures for that purpose. Having these measures defined in the space of stocks we analyze the effects of some extreme phenomena on the geometry of the market. We discuss the hypothesis that collective behavior in period of crises reinforces the structure of correlations between stocks, but that it also may have an opposite effect on clustering by similar economic sectors. Comparing the crises of 1987 and 2001, we discuss why the expansion of the ellipsoid describing the geometry of the distances in the market, which occurs in the latter period, is not homogeneous through sectors. The conclusions from this research identify some of the changes in the structure of the market over the last 30 years.*

## 1. The problem

There is a great deal of empirical research on stock market fluctuations, most of this work has been based in the discovery of patterns both in time and space (cross) correlations between stock returns. When time correlations happen to take place they describe memory patterns, while correlations in space reveal processes of synchronous behavior. Since the behavior of financial markets is considered to be the most remarkable example of complexity in economics, the problem of quantifying cross-correlations in a stock market is important not only from the practical perspective of selecting portfolio strategies, but from the point of view of understanding collective behavior between the elements of a complex system [1].

It is generally accepted that complexity in financial markets is revealed by *(i)* time correlations for series of stocks, considered as units of analysis, *(ii)* space correlations among stocks, and *(iii)* the dynamics of space correlations, which is modified by shocks and crises [2]. Here we also follow an empirical approach and test the hypothesis that space correlations may be extracted from the data itself through the identification of some geometrical relations, namely: the effective dimensionality, the amount of random contributions, the volume and the shape of a market space. Additionally, we add a fourth level of complexity, that of the behavior of the institutional agents contributing to shrink or to grow the market space.

In a previous paper [3], we developed a method for the reconstruction of an economic space. By using a metric related to returns correlation and a stochastic geometry technique, we showed that economic spaces are low-dimensional entities and that this low dimensionality is caused by the small proportion of systematic information present in correlations between stocks. This occurs because the distances between stocks represent both systematic and unsystematic (specific) contributions. The formers are associated to the correlations between the stocks in the market and the latter to individual variances alone. Using our reconstruction method, we found that part of the correlation contribution is indistinguishable from random data, and that the market structure may be represented as a low-dimensional subspace. We also found that the values of the shortest distances between stocks capture maximal information on market synchronization, displaying a completely different behavior depending on the occurrence of bubbles or crashes. Here we discuss whether the synchronous behavior that is observed in periods of crises is uniformly displayed by *i)* market subspaces, *ii)* market periods of expansion and recession and *iii)* sets of stocks sharing the same economic sector.

Some other authors have also been considering the role of economic sectors in the dynamics of stocks. In a recent paper, Marsili [4] showed that economic sectors correspond to clusters of stocks with similar economic dynamics. In the work of Gopikrishman *et all* [5] the authors tried to identify traditional industrial sectors with particular eigenvectors of the correlation matrix of stocks fluctuations. We apply a similar technique although with a different perspective. In our analysis the effective dimension of an economic space may or may not correspond to economic sectors or to any combination of them.

Our empirical data is the set of daily returns of 249 stocks present in S&P500, all those consistent with our time schedule requiring persistence through thirty years (1973 to 2003).

The paper is organized as follows: in section 2 we describe the method we used and the definition of the notions of amount of randomness, volume and shape of an economic space. In section 3, depending on the aspect we are dealing with, the method is applied either to the whole market data or to some selected subset of it. Results are presented in the following way:

- in subsection 3.1 markets of different sizes are investigated. Sub-spacing results confirm the dependency on the effective dimension of: *i)* the market size and *ii)* the number of industrial sectors present in the stocks.
- in subsections 3.2 to 3.4, the entire market (comprising the whole set of stocks) is embed in its effective dimensionality for the purpose of characterizing the amount of randomness present in the market, as well as, the evolution of its volume and shape along the last 30 years.
- subsection 3.5 concerns the analysis of the effects of two major crises (1987 and the period 2001-2002) on the behavior of sector-oriented groups of stocks.

Finally, in section 4 a summary and conclusions are presented.

## 2. Method

Instead of attempting to establish statistical correlations between economic facts and asset returns we follow an empirical approach, testing the hypothesis that it may be possible to extract from the data itself, if not the identification of the variables, at least their geometrical relations. The idea is simply stated in the following terms.

Pick a set of N stocks and their historical data of returns over some time period. From the returns data, using the notion of distance,

$$d_{ij} = \sqrt{2(1 - c_{ij})}$$

as in [6], with $c_{ij}$ being the correlation coefficient of the returns $r_i(t)$,$r_j(t)$; compute the matrix of distances between the N stocks.

The problem now is reduced to an embedding problem where, given a set of distances between points, one asks which is the smallest manifold that contains the set. However, in the distances between assets, computed from their return fluctuations there are systematic and unsystematic contributions. Therefore, to extract further information from the market we have to inquiry whether it is possible to separate these two effects.

## 2.1. The stochastic geometry technique

The following stochastic geometric technique is used for that extraction:

1) From the matrix of distances, compute coordinates for the stocks in an Euclidean space of dimension N-1. The stocks are now represented by a set $X_i$ of points in $R^{N-1}$.
2) To this cloud of points we apply now the standard analysis of reduction of their coordinates to the center of mass and computation of the eigenvalues of the inertial tensor.
3) The same technique is applied to surrogate data (data obtained from independent time permutation for each stock and to random data with the same mean and covariances).

The eigenvalues in (3) are compared with those of (2). The directions for which the eigenvalues are significantly different are now identified as the market systematic variables. We identify the structure that drives the market by means of computing the number of eigenvalues that are clearly different from those obtained from surrogate or random data. The surviving eigenvalues define a subspace $S_e$ of dimension $e$, being $e$ the effective dimension of the economic space.

## 2.2 Volume of a market space

Since the largest $e$ eigenvalues define the effective dimensionality of the economic space $S$, we propose a new measure, the volume ($V$) of $S$, defined as the product of the largest $e$ eigenvalues ($\lambda_1$,$\lambda_2$, ... $\lambda_e$) of $S$.

$$V_e(S) = \prod_{i=1}^{e} \lambda_i$$

## 2.3. Distances in market subspaces

The existence of a market effective dimension allows for the definition of a measure of the shortest average distance $P_e$ in $S$. The value of $P_e$ is computed from the distances $d^{(e)}_{ij}$ between each pair of stocks in $S$, where $d^{(e)}_{ij}$ correspond to the distances $d_{ij}$ restricted to the effective dimension ($e$) of that space.

$$P_e(S) = \frac{1}{N}\sum_{i \neq j}^{N} d_{ij}^{(e)}$$

## 2.4. Market Randomness

In the distances between assets, computed from their return fluctuations, there are systematic and unsystematic contributions. Since the decrease of the eigenvalues of order greater than *e* is indistinguishable from random data, we compute the amount of randomness of a market space *S* as the difference between the distances $P_N$ and $P_e$ of *S*. The value of *RD(S)* provides the amount of random contributions carried in the distances that are not restricted to the effective dimension of the market space.

$$RD(S) = P_N(S) - P_e(S)$$

# 3. Results and Discussion

The measures and technique are applied to the daily returns of 249 stocks present in S&P500, all those consistent with our time schedule requiring persistence through thirty years (1973 to 2003). Depending on the aspect we are dealing with, the method and the measures are applied to both the whole market data or to some selected subset of it.

## 3.1. Subspacing

The calculations have been performed by applying the stochastic geometry technique to actual returns data and to random data with the same mean and variance as the actual data. The first set of actual data consists in the daily returns of 249 stocks. We compute the number of eigenvalues that are clearly different from those obtained from the random data. The ordered eigenvalue distributions obtained in each case are shown in Figure 1.

The plots in Figure 1 represent the largest 25 eigenvalues obtained for the 249-stocks market. The largest 25 eigenvalues are compared to the largest 25 eigenvalues obtained from random data. Given the decrease obtained from the $7^{th}$ eigenvalue, we conclude that the market structure is essentially confined to a *6*-dimensional subspace. Our empirical results show that this dimension captures the structure of the deterministic correlations that are driving the market, and that the remainder of the market space may be considered as being generated by random fluctuations.

In order to investigate if the effective dimension depends on the market size and on the number of industrial sectors present in the stocks, two subsets of 35 stocks are investigated. The first subset comprises 35 firms selected in alphabetical order by the company name from the whole set of 249 stocks. The second subset comprises 13 stocks belonging to the food sector and 22 stocks that belong to the energy sector.

The first two plots in Figure 2 show results for a market comprising 35 stocks that belong to any (among 47) industrial sectors. The last two plots show the same results for a market in which each stock belongs either to the food sector (13 stocks) or to the energy sector (22 stocks). Economic sectors are defined by the first two digits of the Standard Industrial Classification (SIC) codes.

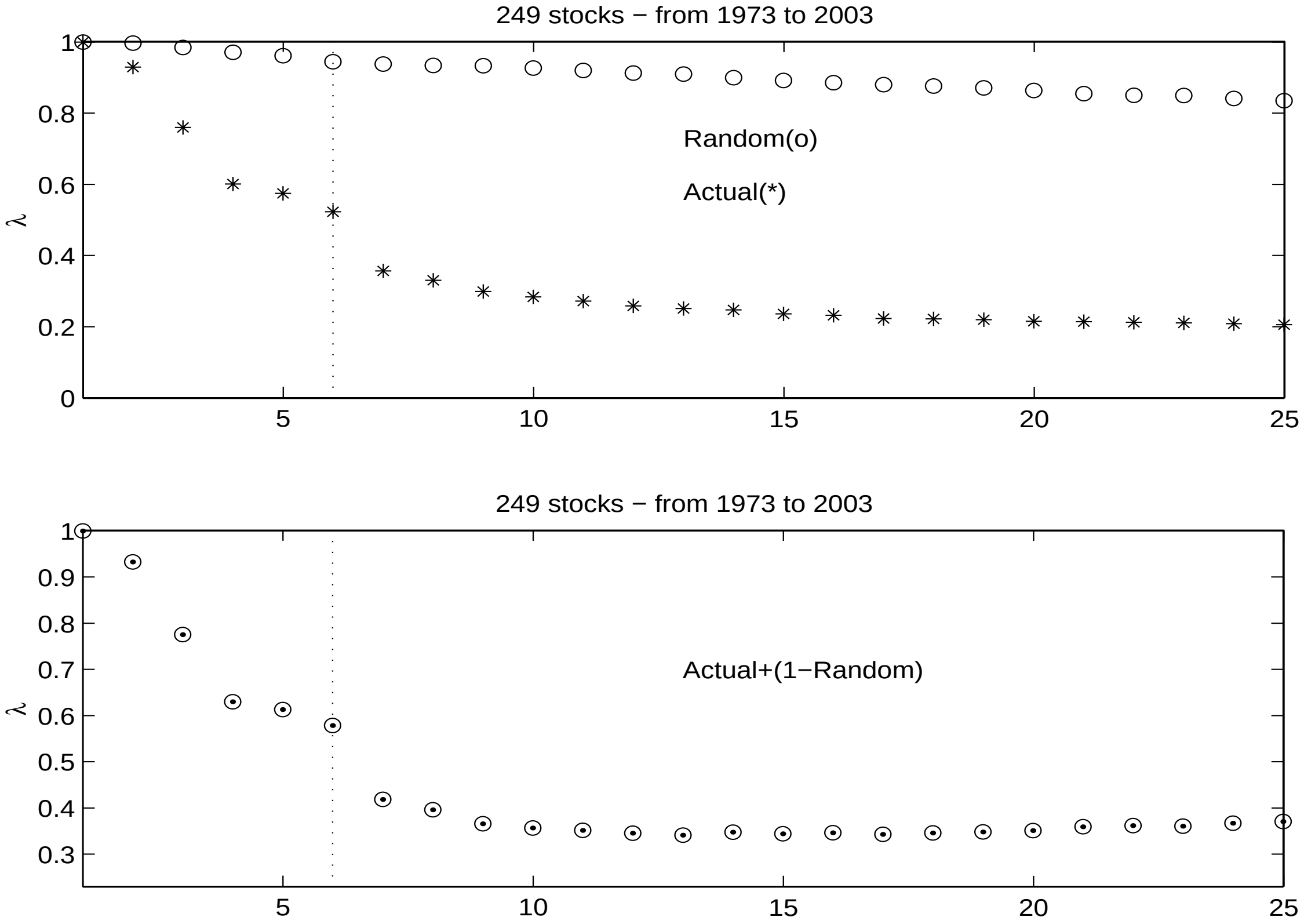

Figure 1: decrease of the largest 25 eigenvalues

As the lower plots in Figure 2 show, when the number of different industrial sectors characterizing the stocks is kept small, the number of surviving eigenvalues is strongly reduced, helping to clarify that economic sectors play a very important role in the definition of the dimensionality of an economic space. The conclusion is that the space effective dimension depends on: *i)* the market size and *ii)* the number of industrial sectors present in the stocks.

## 3.2. Randomness and Clustering

Measures of the average distance and clustering are important statistical parameters used in graph theory to distinguish ordered structures from structures generated at random [7]. A typical random structure is characterized by a short average distance between its elements, whereas in ordered structures - the elements being arranged as in a crystal lattice - the average distance is large. On the other hand, random structures are characterized by a low clustering while the same parameter displays high values for regular frameworks. Another interesting aspect associated to the values of each of those parameters is the characterization of either global or local levels or organization. High clustering usually characterizes a structure that is locally organized, while, in globally organized structures, the average distances between the elements (available paths) are short.

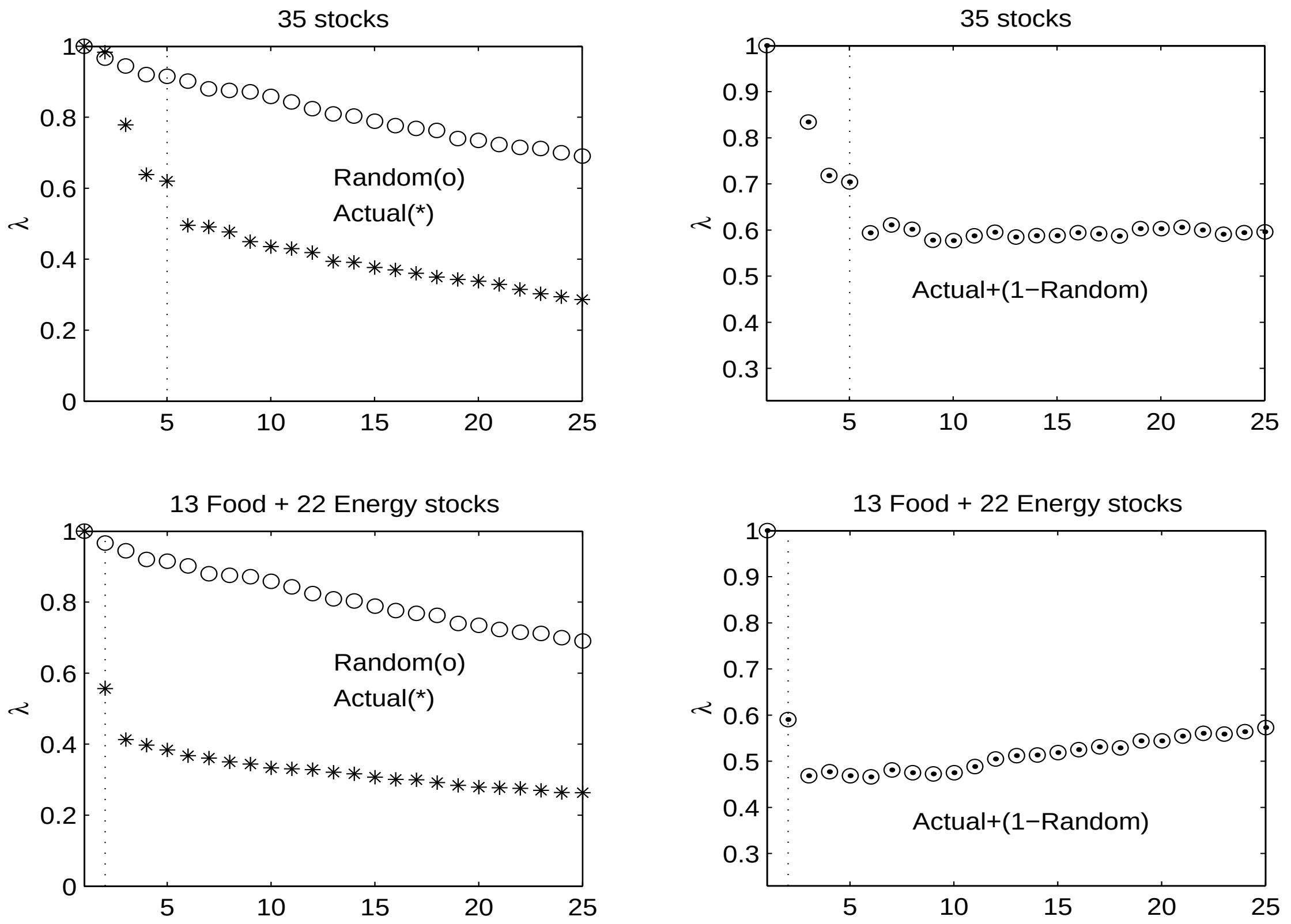

Figure 2: decrease of the largest 25 eigenvalues

In a previous paper [3], we have shown that high clustering is empirically related to periods of market shocks or crises, capturing maximal information on market synchronization in these periods and displaying a completely different behavior in normal periods. The first plot in Figure 3 shows the values of Continuous Clustering [3] computed on a one-year window, for the last 30 years. Years of stock crises (1987 or the period 2001-2002) exhibit the highest clustering values in the last 30 years whereas in normal years, like the period from 1991 to 1995, clustering is very low. The observation of the average values of the distances between stocks (lower plot in Figure 3) computed in the whole space ($R^{248}$) confirms that the market in periods of crises is far from displaying the characteristics of a random structure.

Since the effective dimension of the space is six, we shall measure the distances between stocks in $R^6$. The upper plot in Figure 4 shows that when the distances between stocks are restricted to the effective dimensionality of the market space, the last three years exhibit a quite different behavior when compared to any other year in the set. There are large distances in 2000, 2001 and even 2002, years in which clustering is still very high. An explanation for this disparity may be provided by the computation of the amount of random contribution *RD* in those years.

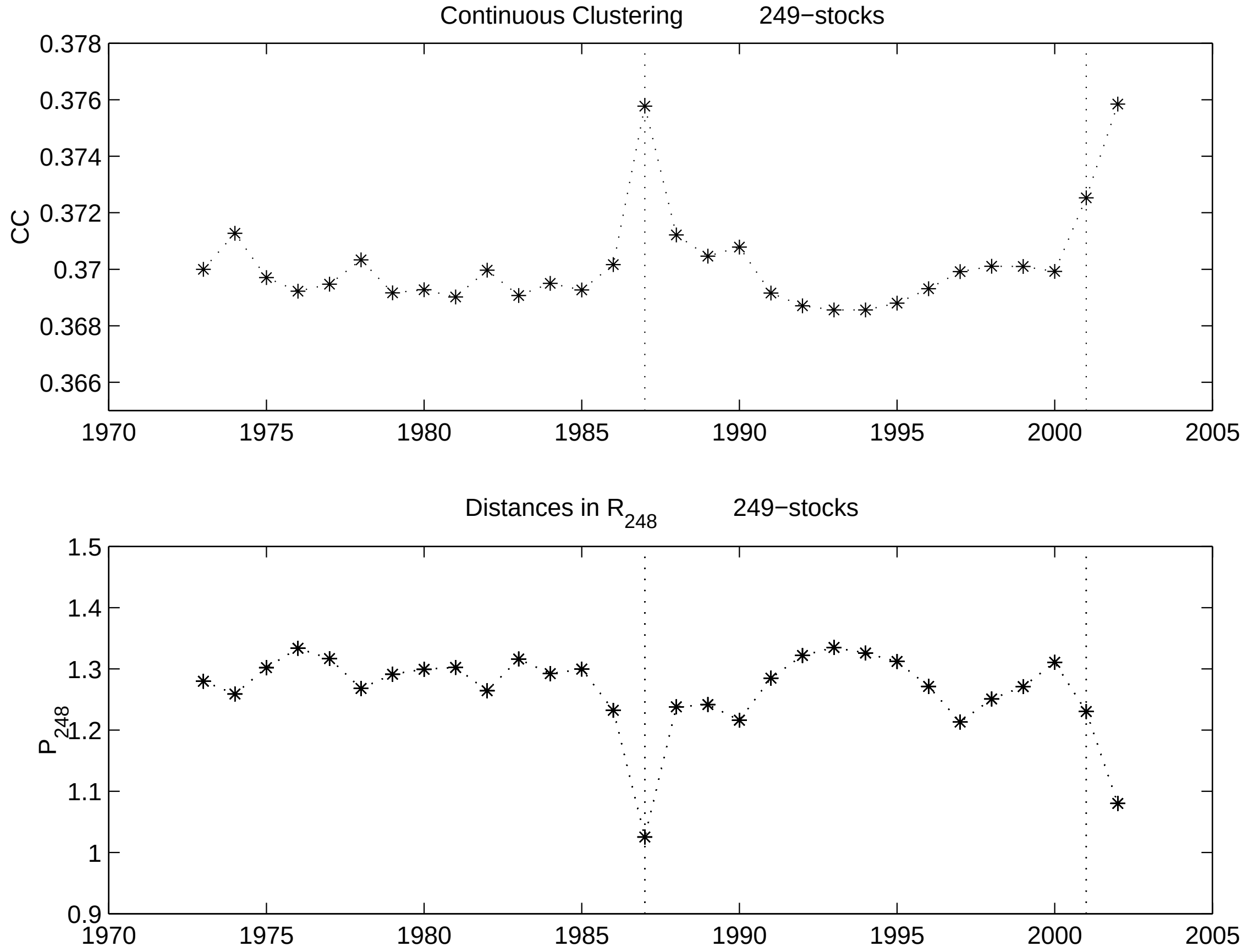

Figure 3: Continuous Clustering and Distance

In computing the values of *RD* for each year between 1973 and 2002 (the lower plot in Figure 4), we see that, as expected, there is a clear correspondence between the fall of the amount of random contribution in each year and the increase of clustering presented in Figure 3. On the other hand, the large distances in 2000, 2001 and even 2002 suggests the emergence of an unusual structure in this last three years: a far from random, highly clustered but dispersal or expanded framework.

## 3.3. Reduction and growth of a market space

We shall focus on the occurrence of either an expanding or a contracting effect in the market space by computing the space volume $V_6$. Figure 5 shows the behavior of $V_6$ from 1973 to 2002. Small circles indicate results obtained from the same calculations applied to random data.

Our empirical results show that $V_6$ displays a quite stable behavior until 1987 but, after the crisis of 1987- which causes a reduction in the volume of the market space in 1988 - there is a slow but persistent expansion until 1993. The information provided within the space volume reinforces the previous idea that in the last three years the market space, in spite of being highly clustered, occupied a large manifold.

Comparing the crises of 1987 and the period 2001-2002, although the years of 1987, 2001 and 2002 share high clustering values, the market in those years seems to behave quite differently with respect to the volume of its dimensional space. The value of $V_6$ in 2001 is the second highest one in the entire set of 30 years, whereas the year of 1987, the moment of the largest stock crisis in the post-1974 period, exhibits a small value for the same coefficient.

As topology is not concerned with the specific shape of objects, the topological nature of clustering measures naturally neglects information on the acquisition of a specific shape. The results obtained for the space volumes suggest that an explanation for the difference between the crises of 1987 and 2001 may be provided by the shape of their corresponding market spaces.

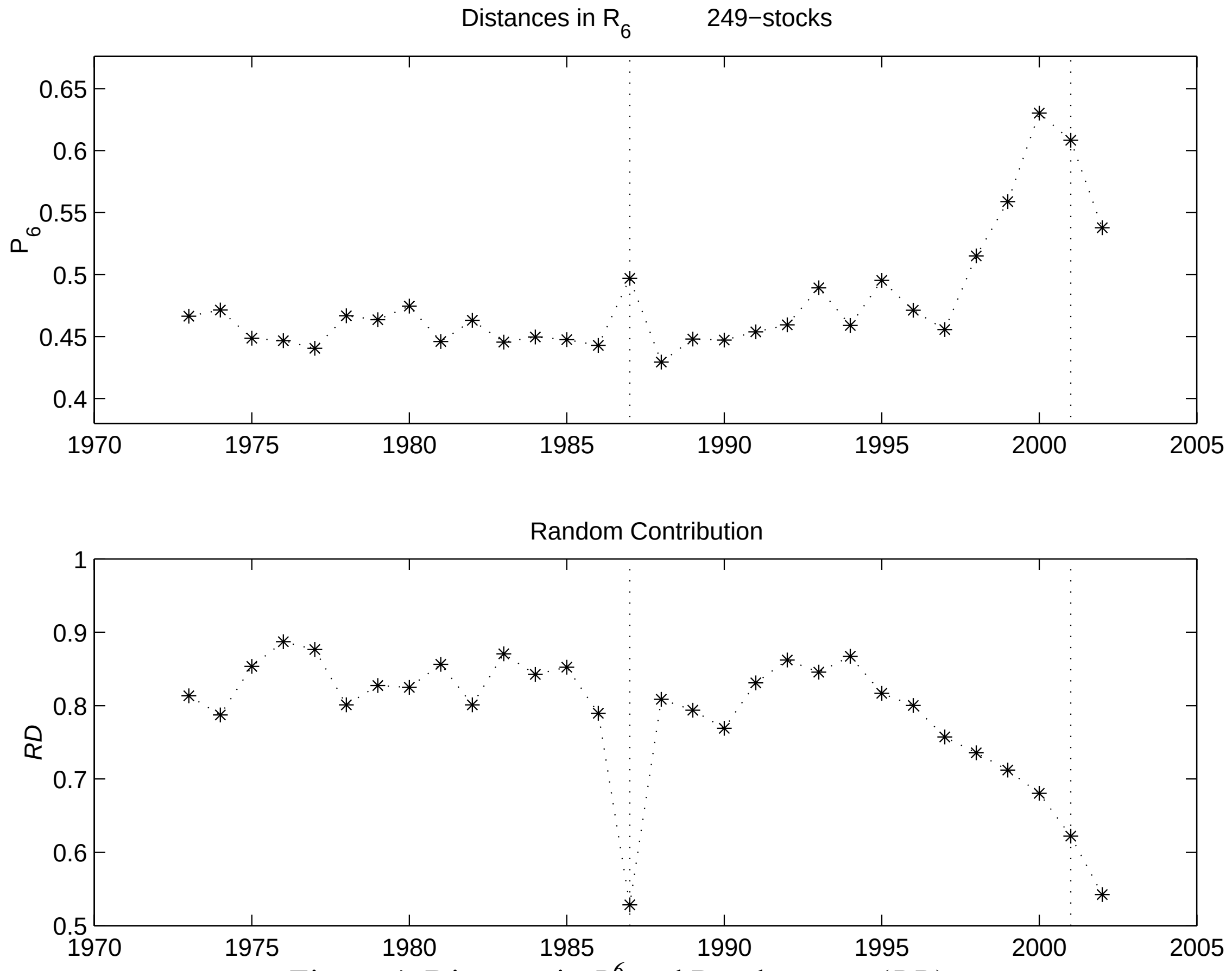

Figure 4: Distance in $R^6$ and Randomness (*RD*)

## 3.4. Observing space shapes

To describe qualitatively how the structure of the subspaces evolves through time, we considered some particular years (1988, 1989, 1999 and 2000) and plotted each yearly batch of 249 stocks in $R^3$. The directions in the plots are those associated to the three largest eigenvalues obtained in each year. Figure 6 shows four plots where each stock is represented by a star (*). Previously we saw that the years of 1987, 2000, 2001 and 2002 share the highest clustering values but exhibit quite different space volumes. Figure 7 shows the shapes of the *3*-dimensional subspaces associated to 1987 and 2001.

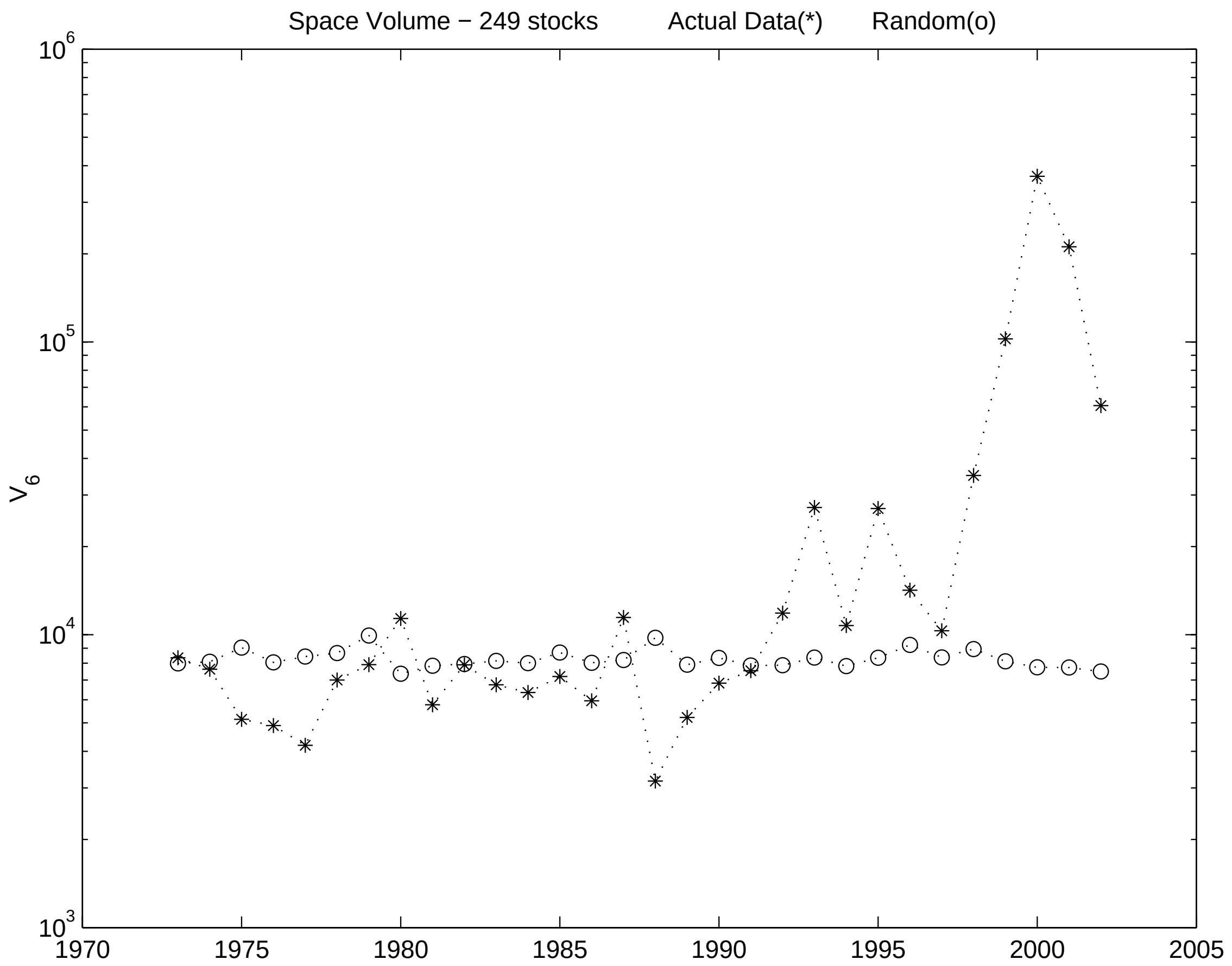

Figure 5: the space volume V (log scale)

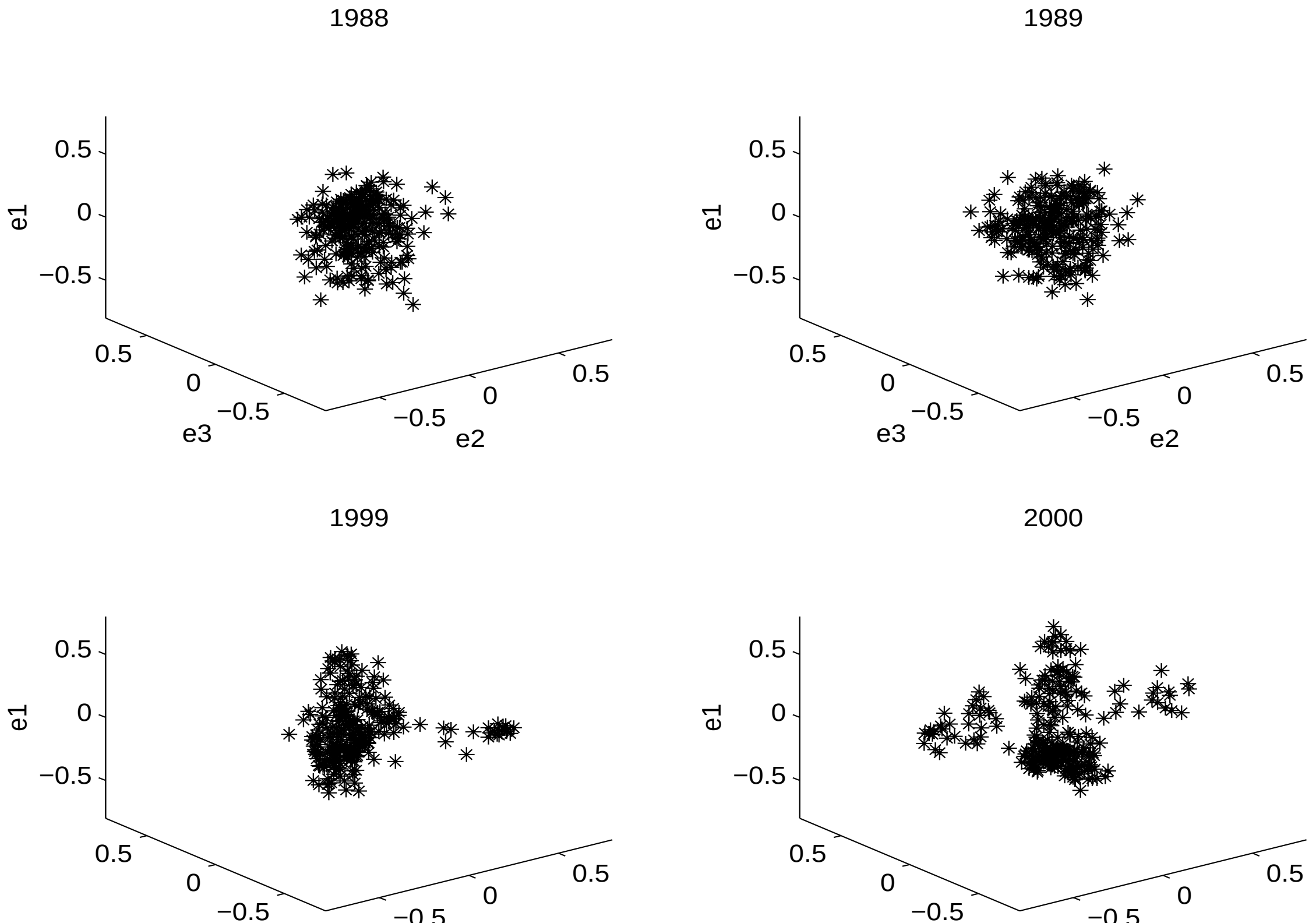

Figure 6: The leading *3*-dimensional subspaces for 1988, 1989, 1999 and 2000

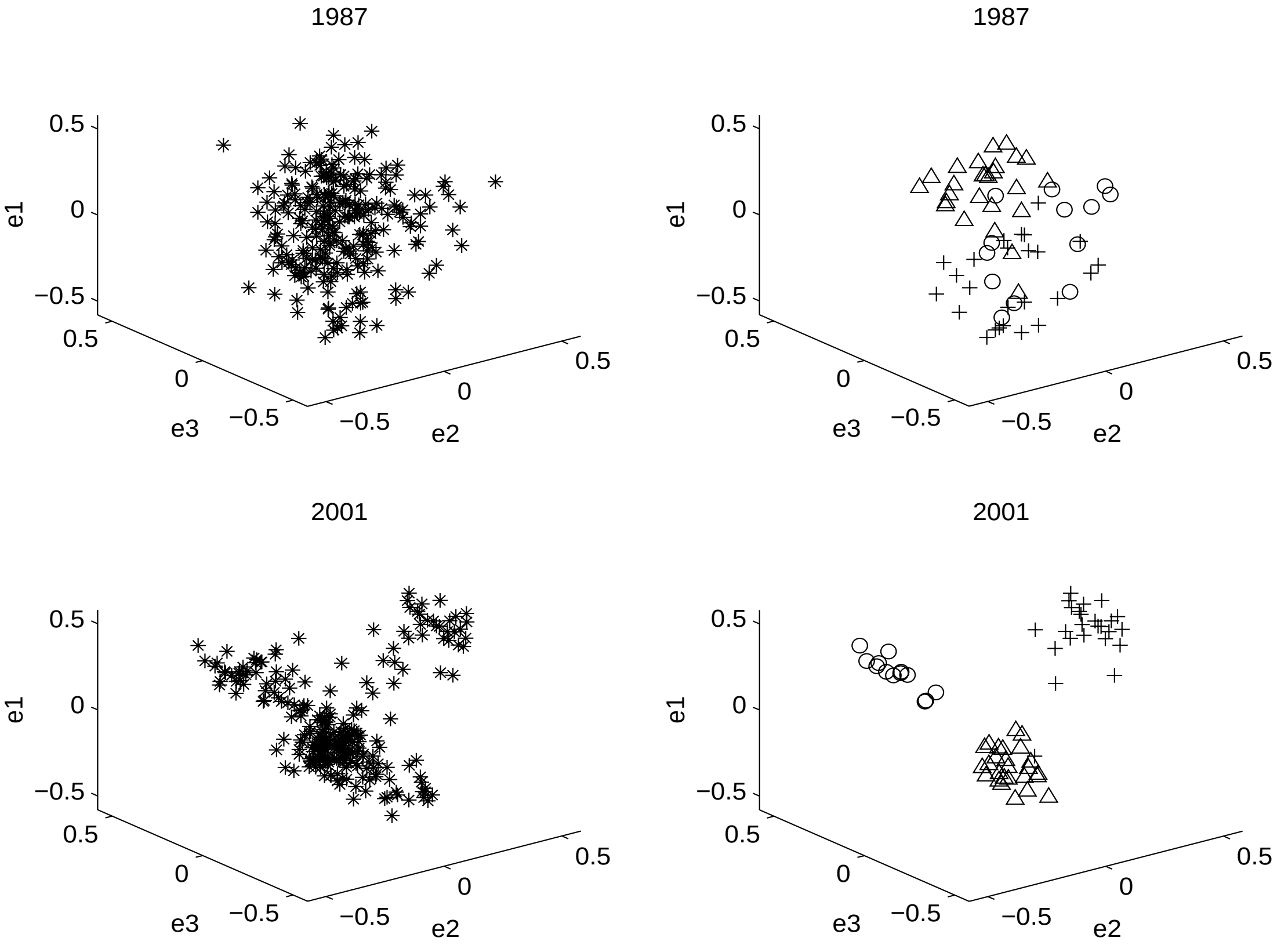

Figure 7: The leading *3*-dimensional subspaces associated to 1987 and 2001: the whole market and some strategic sectors

In order to make the graphs clear, in Figure 7, for each year, two plots were drawn: the plots in the left include the whole market (249 stocks) while the plots in the right include just the stocks belonging to strategic sectors (such as energy (+), food (o), finance or banking (^)). Other sectors were left out in order to avoid overloading the plots.

For the purpose of comparison, the last two plots in Figure 8 include *3*-dimensional subspaces obtained from random data, and, since in random spaces the distribution of the eigenvalues decrease very smoothly, displaying quite similar values for $\lambda_1$, $\lambda_2$ and $\lambda_3$, the random *3*-dimensional spaces approach the spherical shape.

From the observation of the plots we conclude that:

1. The shapes of the space in 1987 and 2001 are remarkably different: while the *3*-dimensional subspace obtained for 2001 is prominent in a particular direction, the subspace obtained for 1987 is closer to the shape of a sphere.
2. In 2000, 2001 and 2002 the prominences in the space shapes seem to correspond to local concentrations of stocks belonging to the same industrial sector.
3. In 1987 stocks in the same industrial sector are mixed and occupy the whole *3*-dimensional subspaces in an indifferentiated way.

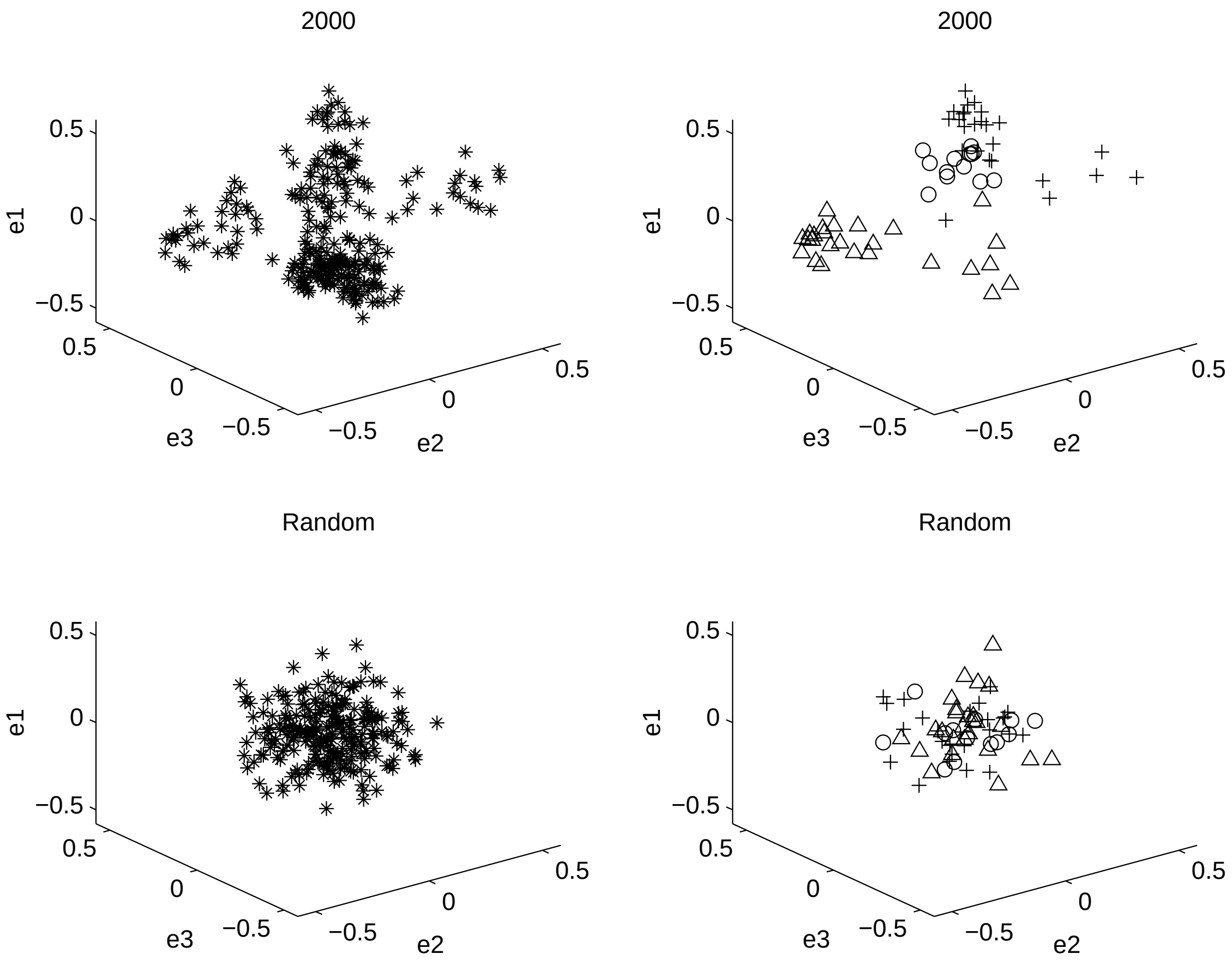

Figure 8: The *3*-dimensional subspaces associated to 2000 and to random data

Since industrial sectors seem to be associated to some remarkable characteristics of *3*-dimensional subspace shapes, in the next section we discuss whether synchronization during crises occurs when we consider either sector-oriented markets or the market as a whole.

## 3.5. Compared Dynamics

When, instead of the whole set of stocks, we consider sub-sets including those stocks sharing the same economic sector and compute the $P_6$ for each of these sub-sets, evidence for some interesting properties emerges. In periods of expansion, sector-oriented sub-sets are characterized by a smaller average distance between stocks. The average behavior of companies belonging to the same economic sector is more synchronous than the behavior of the overall market taken as a whole: tribes of firms act together.

Among others, this effect of synchronization reinforcement can be observed for Food, Banking and Energy sectors. Figure 9 shows the values of the $P_6$ computed for those Energy and Food sectors along the last 30 years. Small circles represent the same values computed for the whole set of stocks. However, in periods of crises like 1973, 1987 and 1997, the average distance of the sector-oriented sub-sets are very similar to the $P_6$ of the whole market, suggesting that in those periods, closeness among stocks that share the same economic sector is lost. The survival effort becomes dominant for the whole market, and it levels the behavior of distinct firms and sectors.

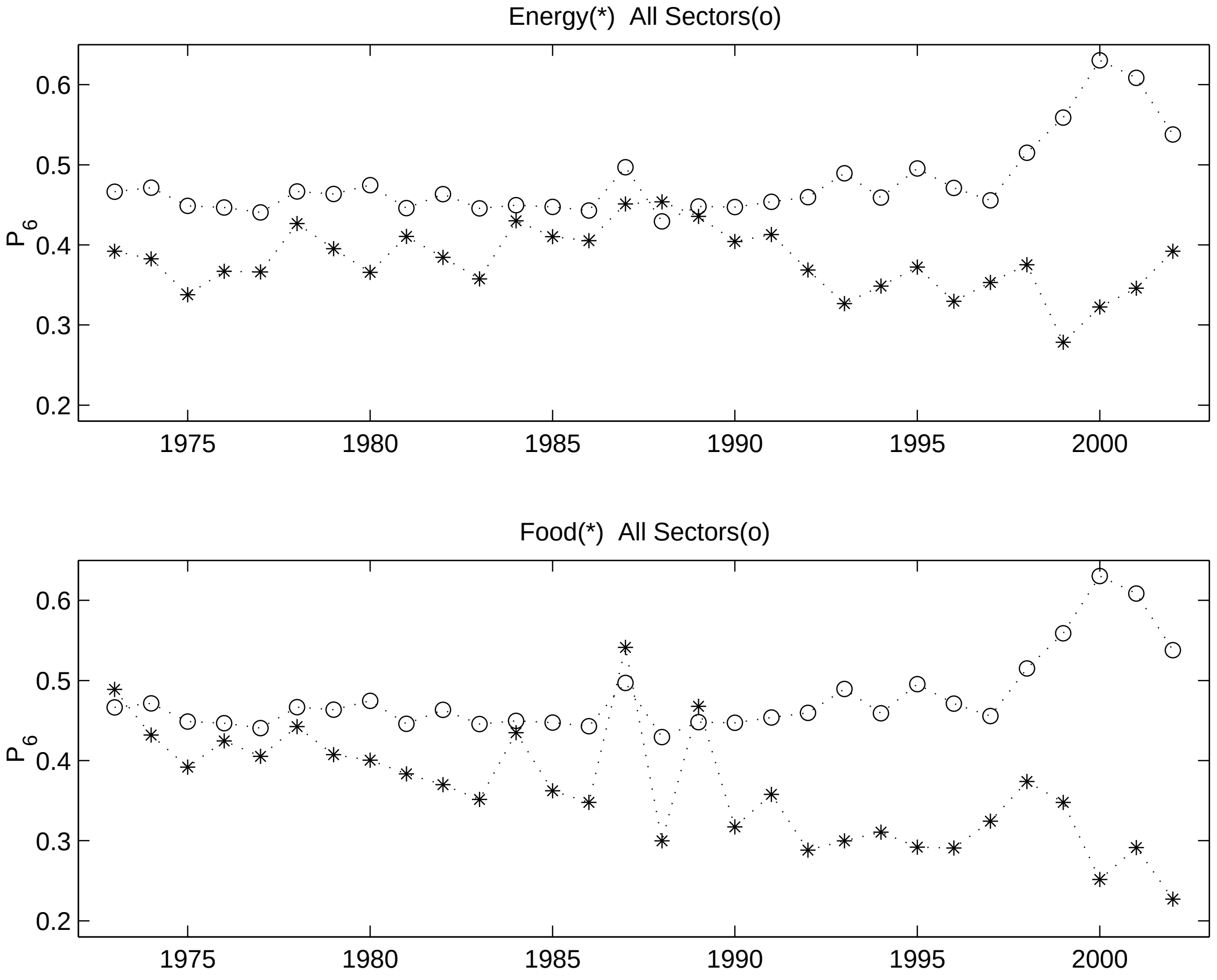

Figure 9: Time pattern of distance for different sectors: Energy and Food

But a remarkable difference characterizes the crises taking place in last three years of our sample: food, banking and energy sectors exhibit $P_6$ values quite smaller than those computed in any other period of time. Moreover, those small values are even more impressive when we see that, as noted in the previous section, the year of 2000 corresponds to a period of expansion, being characterized by very high values of the distances computed for the whole set of stocks. After 2000, a smooth contracting effect on the whole space is detectable, as the stock crisis emerges. However, quite differently from 1987 and 1997, stocks inside banking and food sectors remain closer than ever and a similar proximity defines energy and technological groups – sectors behave as such.

Our explanation for these peculiarities goes as following. Firms with large market power and well placed in strategic sectors, such as those in the energy or the finance and banking clusters, anticipated the effect of disruption of the stock market after the long bubble since 1995. In fact, the "Internet bubble" concentrated expectations in the bandwagon effect of those firms with high profitability, and a virtuous cycle was generated so that they were able to keep a high intensity of investment financed by the stock market itself. Large dividend distributions confirmed the market intuition, even if they were merely self-fulfilling expectations. But this could not last long: after some years this scheme collapsed, and the firms anticipating this evolution took protective measures and reinforced each other's option for defensive action. Furthermore, for some sectors (Energy, Banking, and Food), the 90's and in particular the second half of that period is a period of high divergence in relation to the behavior of the population as a whole: the correlation decreases for all the firms, whereas it increases for these sectors.

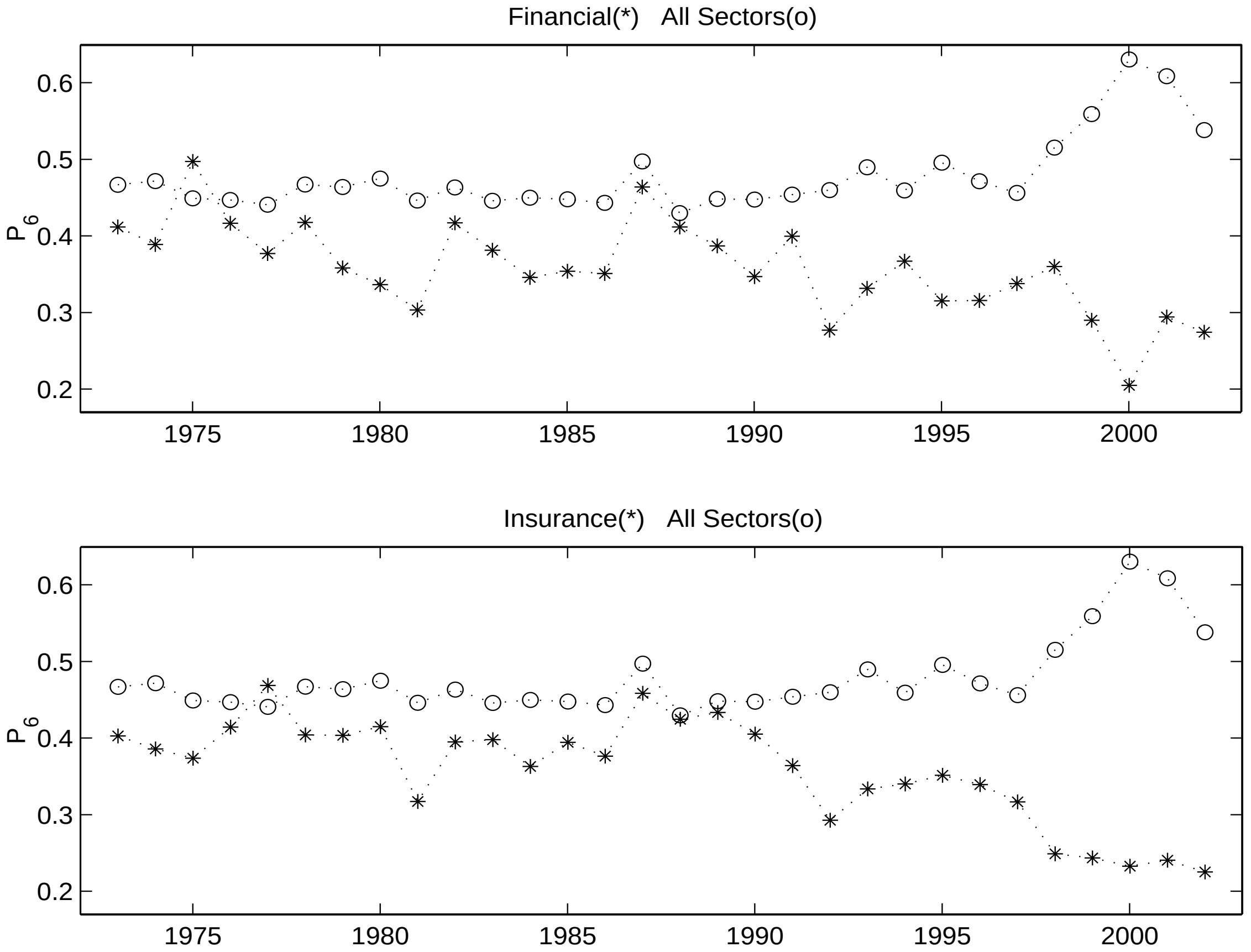

Figure 10: Time pattern of distance for different sectors: Finance and Insurance

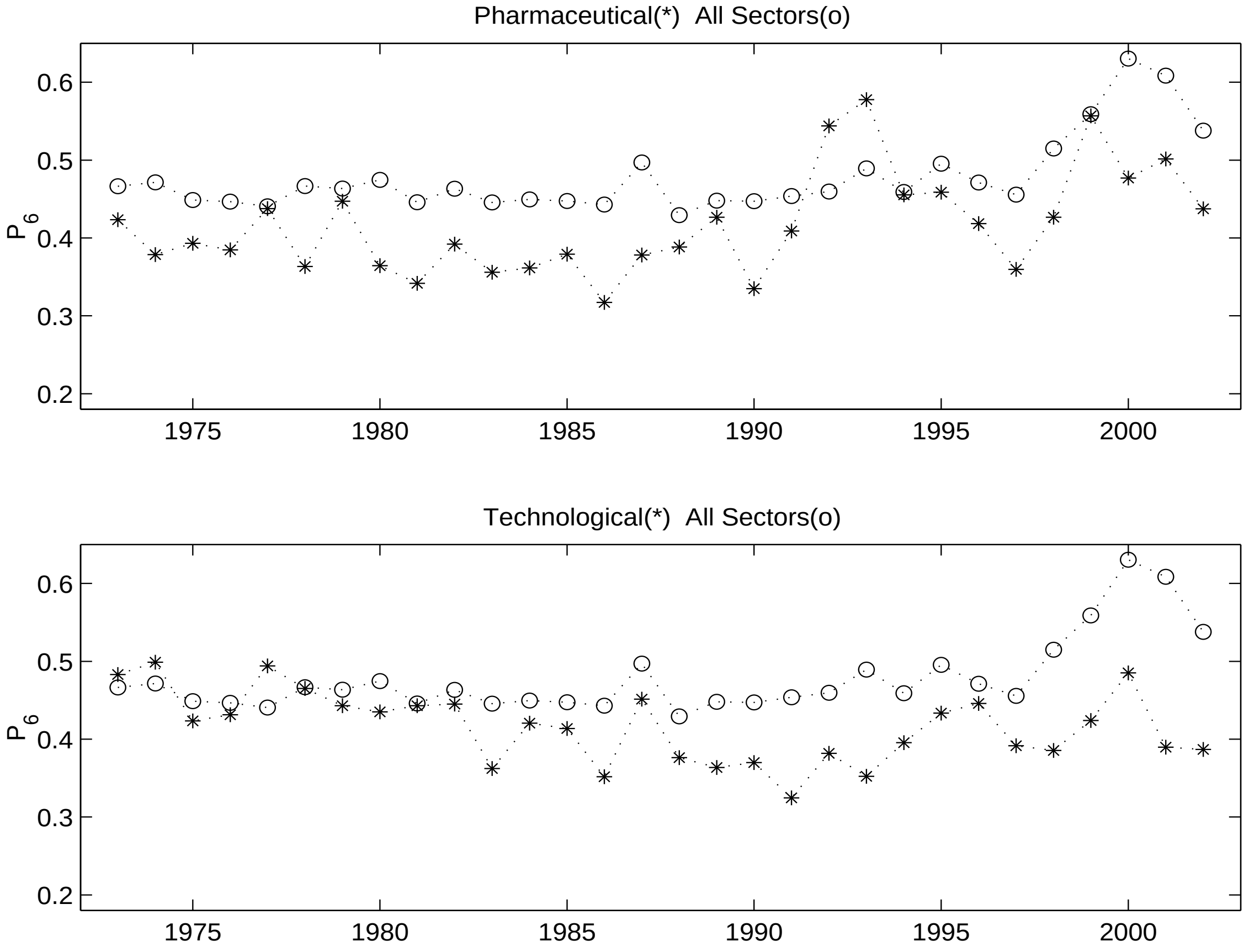

Figure 11: Time pattern of distance for Pharmaceutical and New Technology

In general, during periods of recession, continuous clustering and volatility reach maxima, and synchronous evolution dominates with sectors behaving as one. But the long stock crisis since March 2001 until at least mid-2003 developed differently, and the institutional setup explains such differences.

For the period we are considering, such institutional action is the Greenspan strategy, namely the monetary policy since 1995, strengthening the speculative bubble through the sustained decrease of the interest rate [8]. This latter period is significantly different from the previous crises, since there is a general expansion of the market but a contraction in sectors such as the financial, energy and food firms, as measured in the market space.

## 4. Conclusions

In previous work, the authors discussed a number of concepts and measures designed to provide a metric of the dynamics of stock markets. The method is developed here in order to consider the emergence of collective behavior and to inquiry into the rationale of the reaction to extreme events, the assaults on the normal way of life of sectors defined in economic space of stock markets.

The purpose of the current research is to discuss new methods to measure and to interpret evidence of structure in the evolution of the stock market, namely in order to describe and explain collective behavior in these complex systems. We consider the extreme phenomena, such as crises, both endogenously determined as it may have been the case of that of 1987, and exogenously determined, as that of 2001, as part of its functioning. Consequently, we proceed to describe these events in the framework of the geometrical properties of the ensemble return distribution, and to discuss constrained behavior as part of the emergence of rules of action. In so doing, we were lead to the following main conclusions:

1. Identifying the effective dimensionality of a market space settle the basis for the computation of useful coefficients that carry the systematic contribution present in market data.
2. The behavior of those coefficients show that, differently from what happened in endogenously determined shocks (1987, 1989 and 1997), which lead to close-to-random, tightened volume and spherical shapes, the market space in 2001 is loosened, prominent and shaped by local concentrations of sector-oriented groups of stocks.
3. Although firms generally follow similar collective behavior during crises, different strategies emerge at the same time in each sector: the population reacts as one, but the response of firms differs inside sectors.
4. Since 2001, the lack of homogeneity in the market as a whole associated to synchronous behavior inside sectors corresponds to the emergence of a particular structure which is neither like the low clustered and globally expanded structure that characterizes normal periods nor similar to that high clustered and tightened framework of 1987.
5. When the dynamics of space correlations, modified by shocks and crises, is not homogeneous through sectors there are discrepancies between local (sector-oriented) and global strategies. The topological nature of clustering measures ensures maximal information on synchronization occurring in the entire market but naturally neglects information on the acquisition of any specific shape.
6. Particular shapes emerge from different intra-sector dynamics defining a four level of complexity in the behavior of financial markets.